\documentclass{article}
\usepackage[utf8]{inputenc}
\usepackage{spconf,amssymb,amsmath,graphicx}
\usepackage{multicol,multirow}
\usepackage[hidelinks]{hyperref}
\usepackage{color}
\usepackage[dvipsnames]{xcolor}
\usepackage[noadjust]{cite}  
\usepackage[capitalise]{cleveref}
\usepackage{siunitx}  

\usepackage{caption}
\usepackage{enumitem}\setlist{nosep}  
\usepackage{graphicx}  
\usepackage{xcolor}  
\usepackage{setspace}

\usepackage{anyfontsize}

\graphicspath{ {figures/} }

\makeatletter

\renewcommand{\section}{\@startsection
  {section}%
  {1}%
  {}%
  {-0.2\baselineskip}
  {0.01\baselineskip}
  {}}%

\renewcommand{\subsection}{\@startsection
  {subsection}%
  {2}%
  {}%
  {-0.1\baselineskip}%
  {0.1\baselineskip}%
  {}}%

\renewcommand{\subsubsection}{\@startsection
  {subsubsection}%
  {3}%
  {}%
  {-0.1\baselineskip}%
  {0.1\baselineskip}%
  {}}%

\g@addto@macro\normalsize{%
  \setlength\abovedisplayskip{3pt plus 2pt minus 1pt}
  \setlength\belowdisplayskip{3pt plus 2pt minus 1pt}
  \setlength\abovedisplayshortskip{2pt plus 2pt minus 1pt}
  \setlength\belowdisplayshortskip{2pt plus 2pt minus 1pt}
}

\makeatother

\newcommand{\R}{\mathbb{R}}
\newcommand{\mathsymbol}[1]{{{\scalebox{.6}[1.0]{\fboxsep1pt\fcolorbox{white}{gray!10}{$\mathrm{#1}$}}}}}
\newcommand{\eos}{\mathsymbol{EOS}}
\newcommand{\eoc}{\mathsymbol{EOC}}

\newcommand{\vocab}{\mathcal{A}}

\title{Chunked Attention-based Encoder-Decoder Model for Streaming Speech Recognition}

\name{Mohammad Zeineldeen$^*$\thanks{$*$ denotes equal contribution}, Albert Zeyer$^*$, Ralf Schlüter, Hermann Ney}
\address{
Machine Learning and Human Language Technology, Computer Science Department, \\
RWTH Aachen University, Germany \\
AppTek GmbH, Germany 
}

\begin{document}
%
\maketitle


\begin{abstract}
We study a streamable attention-based encoder-decoder model
in which either the decoder, or both the encoder and decoder,
operate on pre-defined, fixed-size windows called chunks.
A special end-of-chunk (EOC) symbol advances from one chunk to the next chunk,
effectively replacing the conventional end-of-sequence symbol.
This modification,
while minor,
situates our model as equivalent to a transducer model that operates on chunks instead of frames,
where EOC corresponds to the blank symbol.
We further explore the remaining differences between a standard transducer and our model.
Additionally, we examine relevant aspects such as
long-form speech generalization, beam size, and length normalization.
Through experiments on Librispeech and TED-LIUM-v2, and by concatenating consecutive sequences for long-form trials,
we find that our streamable model maintains competitive performance compared to the non-streamable variant and generalizes very well to long-form speech.
\end{abstract}

\section{Introduction \& Related Work}

Among the potential streaming models,
there are the traditional HMM \cite{Bourlard1994HMM},
CTC \cite{Graves2006CTC}
and more recently transducer \cite{graves2012rnnt}.
While many streamable attention-based encoder-decoder (AED) models were proposed
\cite{chorowski2015attention,chiu2018mocha,Hsiao2020OnlineAED,Tsunoo2021Blockwise,zeyer2022:segmental-attention},
they are too complicated, relying on too much heuristics and not being robust enough in comparison to the transducer
\cite{prabhavalkar2023endtoend}.

Here we show, how a seemingly very simple modification
makes the AED model streamable and turns out to be very robust and competitive, specifically on long-form speech,
in contrast to many other AED and transducer models
\cite{Narayanan2019LongForm,Chiu2019LongForm,zeyer2020:transducer,zeyer2022:segmental-attention,prabhavalkar2023endtoend}.
Interestingly, the small modification leads to an equivalence to transducer models,
and we study the exact modeling differences.

We use \emph{chunking}
as the core mechanism
for both the encoder
and cross-attention in the decoder.
This means that we take out chunks (windows) of \emph{fixed width}
and \emph{fixed step sizes (striding)}.
The static step size implies
that we have a variable number of labels per chunk.
The static sizes in the encoder also allow for efficient processing in training and recognition,
more efficient than causal self-attention
and also performing better.

Related to chunkwise processing
is the operation on segments with variable boundaries
in segmental attention models \cite{zeyer2022:segmental-attention},
or on fixed-size windows at variable positions
\cite{zeyer2021:latent-attention}.
Having variable positions or segment boundaries
allows to use a single label
per window or segment.
In contrast, using fixed-size chunks at fixed positions
implies that we have a variable number of labels per chunk.
Further, we can use the same chunking in the encoder,
with the big advantage that we can parallelize the training computation
in the encoder independent of the alignment.

Similar chunking in the decoder has been done in \cite{Jaitly2016ChunkTransducer,Sainath2018ChunkedTransducer,Tian2020SyncTrafo,Wilken2020SimultaneousTranslation,radford2023whisper,Zhang2023GoogleUSM}
and similar chunking in the encoder has been done in
\cite{zeyer16:onlinebidir,Dong2019ChunkHopping,Tsunoo2019BlockProc,Zhang2020UnifiedStreaming,Tsunoo2021Blockwise,emformer,Chen2021StreamingTrafo,An2022cuside,Weninger2022DualMode,Swietojanski2023AttMasking,Gulzar2023MiniStreamer}.
There are also other approaches to make self-attention in the encoder streamable
\cite{Wang2020ScoutNet,moritz21_interspeech,prabhavalkar2023endtoend}.





\section{Global AED Model}


Our baseline is the
standard global attention-based encoder-decoder (AED) model \cite{bahdanau2015nmt}
adapted for speech recognition
\cite{chorowski2015attention,chan2016las,%
park2019specaugment,tuske2020swbatt}.
We use a Conformer-based \emph{encoder} \cite{gulati_conformer}.
The model operates on a sequence of audio feature frames $x_{1:T} \in \R^{T \times D}$ (10ms resolution) of length $T$ as input
and encodes it as a sequence
\[ h_{1:T'} = \operatorname{GlobalEncoder}(x_{1:T}) \in \R^{T' \times D_{\textrm{enc}}} \]
of length $T'$ and encoder feature dimension $D_{\textrm{enc}}$.
The encoder has a convolutional frontend with striding in time
which downsamples the input by a factor of 6.
Thus, the encoder outputs a frame every 60ms
and $T' = \lceil \frac{T}{6} \rceil$.

The probability of the output label sequence $a_{1:S} \in \vocab^S$ 
given the encoder output sequence $h_{1:T}$ 
is defined as
\begin{align*}
p(a_{1:S} \mid h_{1:T'})
&=
\prod_{s=1}^{S}
p(a_s \mid a_{1:s-1}, h_{1:T'}).
\end{align*}
We have $a_S = \eos$ to mark the end of the sequence (EOS),
which implicitly models the probability of the sequence length.
This part of the model is called the \emph{decoder}.
The decoder uses \emph{global attention} on $h_{1:T}$ per output step $s$.
The main and sole difference
of the global decoder vs.~the chunked decoder
is global attention vs.~chunked attention.
The decoder is defined below.

\section{Chunked AED Model}

\begin{figure}
\centering
\includegraphics[width=0.9\columnwidth]{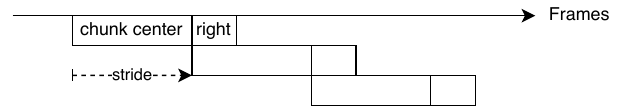}
\caption{\it Chunking on input frames $x_{1:T}$ with chunk center size $T_w$, right context $T_r$ and stride $T_s$, where we have $T_s = T_w$.}
\label{fig:chunking}
\end{figure}

As visualized in \cref{fig:chunking},
we extract strided windows called `chunks'
with chunk size $T_w$ and stride $T_s$.
For input $x_{1:T} = (x_1, \dots, x_T) \in \R^{T \times D}$,
we get the chunks $x'_{1:K, 1:T_w} \in \R^{K \times T_w \times D}$
with
$x'_{k,1:T_w} \in \R^{T_w \times D}$
for chunk index $k \in \{1,\dots,K\}$
with $K = \lceil \frac{T}{T_s} \rceil$,
where
\[x'_{k,t} = x_{(k-1) \cdot T_s + t} \in \R^D,
\quad t \in \{1,\dots,T_w\} . \]
Additionally, we might extend the chunk size by $T_r$ more frames
to get some extended right context.

For the streaming model, the chunking is applied directly on the input
(e.g.~log mel features every 10ms),
and then a variant of the Conformer encoder works on the chunks
$x'_{1:K, 1:T_w}$
and calculates the encoder output
\[ h'_{1:K, 1:T'_w} = \operatorname{ChunkedEncoder}(x'_{1:K, 1:T_w}) \in \R^{K \times T'_w \times D_{\textrm{enc}}} \]
where $T'_w = \lceil \frac{T_w}{6} \rceil$.

For comparison, we also use a standard Conformer
with global attention applied on the whole input
\[ h_{1:T'} = \operatorname{GlobalEncoder}(x_{1:T}) \in \R^{T' \times D_{\textrm{enc}}} \]
and apply chunking on the encoder output $h_{1:T'}$
such that we get the chunked encoder output
$h'_{1:K, 1:T'_w}$.

\subsection{Streamable Chunked Encoder}

\begin{figure}
\centering
\includegraphics[width=0.6\columnwidth]{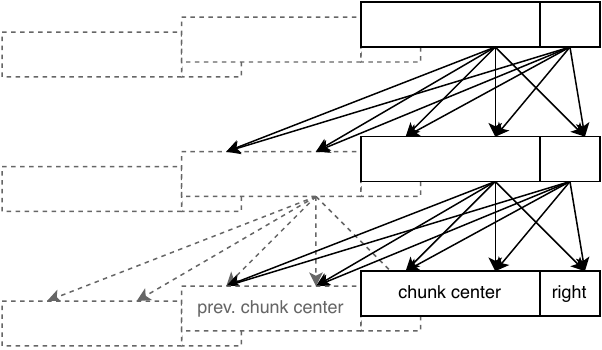}
\caption{\it Chunked self-attention in the encoder.}
\label{fig:chunk-self-att}
\end{figure}

Our starting point is the standard Conformer,
operating on chunks instead of the whole sequence,
i.e.~operating on $x_{k,1:T_w}$ for every chunk index $k$.
The self-attention is calculated per chunk,
i.e.~both the chunk center and right context frames,
and attends to all frames within the chunk,
and additionally to the previous chunk,
as can be seen in \cref{fig:chunk-self-att}.
Thus it is non-causal within the chunk,
just like the convolution.
The decoder cross-attention will afterwards
only access the chunk center frames,
thus we expect that the chunk center covers
the labels for this chunk.
The future lookahead via the right context frames
does not accumulate over multiple layers,
unlike the history context,
where we access the previous chunk,
thus the history context does accumulate over multiple layers.
This also explains why
we don't need to have any additional left context frames within the chunk.

In training, we can calculate all chunks in parallel,
and the self-attention calculation per chunk
is more efficient compared to the global self-attention.
We only get a small overhead due to the overlap of the chunk via the right context frames.

Note that this is mathematically
equivalent to the same kind of look-ahead context leaking avoidance as in the
Emformer \cite{emformer}
and dual causal/non-causal self-attention \cite{moritz21_interspeech}.

\subsection{Streamable Chunked Decoder}

\begin{figure}
\centering
\includegraphics[width=0.55\columnwidth]{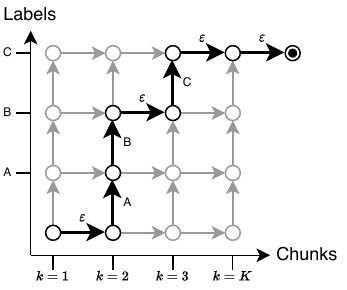}
\vspace*{-2pt}
\caption{\it Possible transition sequences $a_{1:K+N}$ for non-EOC label sequence $ABC$ with length $N=3$ and $K=4$ chunks,
where $\varepsilon$ is the end-of-chunk (EOC) symbol.}
\label{fig:lattice}
\end{figure}

In the output vocabulary $\vocab$, we replace the $\eos$
by a new special end-of-chunk (EOC) symbol $\eoc$.
We start with the first chunk ($k=1$),
and once we get $\eoc$,
we advance to the next chunk ($k' = k + 1$).
The decoder is exactly like in the global AED model,
except that the global attention is replaced by attention on the current chunk.
The possible transitions can be seen in \cref{fig:lattice}.

The probability to emit the next label $a_s \in \vocab$
is estimated using a LSTM \cite{hochreiter1997lstm} with zoneout \cite{krueger2017zoneout}
and MLP cross-attention \cite{bahdanau2015nmt}
to the current chunk of the encoder:
\begin{align*}
p(a_s \mid ...)
&= (\operatorname{softmax} \circ \operatorname{Linear} \circ \operatorname{maxout} \circ \operatorname{Linear}) \big(
g_s, c_s \big)
\\
g_s &= \operatorname{ZoneoutLSTM}(c_{1:s-1}, a_{1:s-1}) \\
c_s
&= \sum_{t=1}^{T'_w}
\alpha_{s,t}
\cdot h'_{k_s,t}
\in \R^{D_{\textrm{enc}}}
\\
\alpha_{s,t}
&=
\frac{\exp (e_{s,t})}{\sum_{\tau=1}^{T'_w} \exp (e_{s,\tau})}
\in \R, \quad t \in \{1,\dots,T'_w\}
\\
e_{s,t}
&=
(\operatorname{Linear} \circ \tanh \circ \operatorname{Linear}) \big(
g_s, h_t \big) \in \R ,
\end{align*}
and the current chunk index $k_s$ is defined as
\[
k_s = \begin{cases}
k_{s-1} + 1, & a_{s-1} = \eoc \\
k_{s-1}, & a_{s-1} \ne \eoc
\end{cases}
\]
and initially $k_1 = 1$.
The sequence is ended when we reach $k_s = K$ and $a_s = \eoc$.
The attention weights here are only calculated inside the current chunk.
Further, we do not use attention weight feedback.
Otherwise the model is exactly the same as the global attention decoder,
to allow for direct comparisons, and also to import model parameters.

We realize that the chunked decoder is equivalent to a transducer model \cite{graves2012rnnt,zeyer2020:transducer},
where $\eoc$ behaves exactly like the blank symbol,
and we iterate over chunks instead of frames,
which is like a higher downsampling rate.
A similar observation for a similar model
has been made in \cite{Tian2020SyncTrafo}.
The main difference is the cross-attention
and the decoder LSTM dependence on the encoder output.
Note that this is a different kind of equivalence
compared to \cite{zhou:segmental-transducer:interspeech2021},
where a segmental model is rewritten in a framewise manner.

\subsection{Training}

We create a chunkwise alignment from
an existing framewise alignment,
then add the EOC labels,
and train with labelwise cross-entropy,
just like the standard AED training.
This is different to the standard transducer training,
which performs a full sum over all alignment paths.
The standard transducer training criterion
cannot be applied easily here due to the alignment label dependencies \cite{zeyer2020:transducer,prabhavalkar2023endtoend}.

\subsection{Beam Search}

We perform alignment-synchronous search,
meaning that in each step,
all hypotheses have the same number of labels,
including $\eoc$.
It is exactly the same as the alignment-synchronous transducer search
\cite{Saon2020AlignSync,zeyer2020:transducer}.


For the very best results, we make use of an external language model (LM)
and perform internal language model (ILM) prior correction \cite{zeineldeen_ilm}.
Note that the chunked AED model has the EOC label (blank label) instead of the EOS label.
We use the scores
\[
    P(a_s \vert ...) = 
    \begin{cases}
            P_\textrm{AED}^{\alpha}(a_s \vert ...) \\ {}\quad\cdot P^{\beta}_{\textrm{LM}}(a_s \vert ...) \\ {}\quad\cdot P^{-\lambda}_{\textrm{ILM}}(a_s \vert ...), & a_s \neq \eoc \\
		  P_\textrm{AED}(\eoc \vert ...), & a_s = \eoc, k < K \\
            P_\textrm{AED}^{\alpha}(\eoc \vert ...) \\ {}\quad\cdot P^{\beta}_{\textrm{LM}}(\eos \vert ...) \\ {}\quad\cdot P^{-\lambda}_{\textrm{ILM}}(\eos \vert ...), & a_s = \eoc, k=K \
	\end{cases}
\]
where $\alpha$, $\beta$, and $\lambda$ are tuned scales and we set $\alpha = 1 - \beta$.
$P_\textrm{LM}(\eos \vert ...)$ and $P_\textrm{ILM}(\eos \vert ...)$ are set to 0 for $k < K$ and then renormalized.
This is very similar to the EOS handling for transducer models with ILM prior correction
\cite{zeyer2021:transducer-librispeech}
except that our ILM also has EOS
and the renormalization.
We use Mini-LSTM ILM method \cite{zeineldeen_ilm} that is trained on the train transcription
labels with $\eos$ same as LM training data.

\section{Experiments}

We conduct experiments on LibriSpeech 960h \cite{libri_corpus} 
and TED-LIUM-v2 200h \cite{ted2_corpus}
using BPE labels \cite{bpe_paper}.
We use RETURNN \cite{zeyer2018:returnn}
based on TensorFlow \cite{tensorflow}.
All code including full recipes are online%
\footnote{\tiny\url{https://github.com/rwth-i6/returnn-experiments/tree/master/2023-chunked-aed}}.

We train the global AED model for 100 epochs
using a single consumer GPU.
We apply on-the-fly speed perturbation and SpecAugment \cite{park2019specaugment}.
The encoder consists of 12 Conformer layers with 
512 model dims.~and 
decoder LSTM has 1024 dims.
We use an aux.~CTC loss \cite{Hori2017CtcAtt}
on top of encoder output
for better training convergence
and for the alignments.
%

To train the chunked AED models, we extract time-synchronous alignments
from the jointly trained CTC model with disallowed label loop.
Then, we convert such alignment into a chunk-synchronous one and use that
as targets for cross-entropy training.
We initialize all parameters using the best checkpoint of the global AED model 
and train for 15-30 epochs.





\subsection{Chunked Decoder}

\begin{table}
    \caption{
    \it WERs [\%],
    studying \textbf{chunked decoder}
    with different \textbf{chunk sizes} with no overlap
    when using \textbf{global encoder}.
    $\infty$ means global decoder.
    The frame rate of $h$ is 60 ms.
    }
    \label{tab:chunked-decoder}
    \centering
    \begin{tabular}{|S[table-format=2]|S[table-format=1.2]||c|c||c|c|}
         \hline
         \multicolumn{2}{|c||}{Chunk size} & \multicolumn{2}{c||}{TED-v2} & \multicolumn{2}{c|}{LibriSpeech} \\ \cline{1-6}
         {$T'_w$} & {Sec.} & dev & test & dev-other & test-other \\
         \hline\hline
         1 & 0.06 & 7.5 & 7.3 & 5.8 & 6.0 \\ \hline
         5 & 0.3 & 7.3 & 7.1 & 5.7 & 5.9 \\ \hline
         10 & 0.6 & 7.3 & 6.9 & 5.7 & 5.7 \\ \hline
         25 & 1.5 & 7.4 & 6.9 & 5.6 & 5.7 \\ \hline
         \hline
         \multicolumn{2}{|c||}{$\infty$} & 7.4 & 6.9 & 5.6 & 5.7 \\ 
        \hline
    \end{tabular}
\end{table}

First,
we investigate the effect of chunking only in the decoder,
i.e.~chunking the output $h$ of the global encoder.
Results on TED-LIUM-v2 and LibriSpeech are shown in \cref{tab:chunked-decoder}.
We can observe that we are able to achieve same WERs as the global AED 
model even with small chunk sizes.


\subsection{Chunked Encoder-Decoder}

\begin{table}[t]
    \centering
    \caption{
    \it For \textbf{chunked AED}, effect on WERs[\%] for \textbf{carry-over} history context, center \textbf{chunk size} $T_w$, \textbf{lookahead} future context $T_r$.
    All sizes are in seconds.
    }
    \label{tab:chunked-enc-dec}
    \setlength{\tabcolsep}{3pt}
    \begin{tabular}{|c|c|c||c|c||c|c|}
        \hline
        \multirow{2}{*}{\shortstack{Carry-\\over}} & \multirow{2}{*}{\shortstack{Chunk\\size}} & \multirow{2}{*}{\shortstack{Look-\\ahead}} & \multicolumn{2}{c||}{TED-v2} & \multicolumn{2}{c|}{LibriSpeech} \\ \cline{4-7}        
         & & & dev & test & dev-o. & test-o. \\ \hline
        \hline
        \multirow{3}{*}{2.4} & \multirow{3}{*}{0.6} & 0.3 & 8.2 & 7.6 & 7.2 & 7.4 \\ \cline{3-7}
            & & 0.6 & 7.9 & 7.4 & 6.8 & 6.8 \\ \cline{3-7}
            & & 0.9 & 7.7 & \textbf{7.1} & 6.6 & 6.7 \\ \hline
        \hline
        0\phantom{.0}   & \multirow{4}{*}{1.2} & \multirow{4}{*}{0.3} & 8.6 & 8.0 & 7.1 & 7.0 \\ \cline{1-1} \cline{4-7}
        1.2 & & & 7.9 & 7.3 & 6.8 & 6.8 \\ \cline{1-1} \cline{4-7}
        2.4 & & & 7.7 & 7.3 & 6.7 & 6.7 \\ \cline{1-1} \cline{4-7}
        3.6 & & & 7.7 & 7.3 & 6.7 & 6.7 \\ \hline 
        \hline
        \multirow{3}{*}{2.4} & \multirow{3}{*}{1.2} & 0\phantom{.0} & 10.2 & 9.7 & 7.8 & 7.8 \\ \cline{3-7}
        & & 0.6 & 7.8 & 7.2 & 6.5 & 6.6 \\ \cline{3-7}
        & & 0.9 & \textbf{7.5} & \textbf{7.1} & \textbf{6.2} & 6.3 \\ \hline
        \hline
        3.0 & 1.5 & \multirow{2}{*}{0.3} & 7.7 & 7.3 & 6.3 & 6.3 \\ \cline{1-2} \cline{4-7}
        3.6 & 1.8 & & \textbf{7.5} & \textbf{7.1} & \textbf{6.2} & \textbf{6.2} \\ \hline
        \hline
        \multicolumn{3}{|c||}{$\infty$} & 7.4 & 6.9 & 5.6 & 5.7 \\ \hline
    \end{tabular}
\end{table}

\Cref{tab:chunked-enc-dec} shows WER results of the chunked AED model.
We observe that carrying over left context yields improvement,
where 2.4 seconds is enough.
In addition, using future lookahead gives good improvements in all cases.
The chunked AED model with a total chunk size and lookahead
of 2.1 seconds achieves a WER of $7.1\%$ and $6.2\%$ on TED-LIUM-v2
and LibriSpeech test sets respectively,
a relative increase in WER of $4\%$ and $9\%$ compared to global AED model.

\subsection{Latency}

\begin{table}[t]
    \centering
    \caption{
    \it Word emit \textbf{latency} for chunked AED model on TED-LIUM-v2 dev dataset. All timing values are in seconds.
    }
    \label{tab:latency}
    \setlength{\tabcolsep}{3pt}
    \begin{tabular}{|c|c|c||c|c|c||c|}
        \hline
         \multirow{2}{*}{\shortstack{Carry-\\over}} & \multirow{2}{*}{\shortstack{Chunk\\size}} & \multirow{2}{*}{\shortstack{Look-\\ahead}} & \multicolumn{3}{c||}{Latency} & WER [\%] \\ \cline{4-7}
         &&& $\% 50^{\textrm{th}}$ & $\% 95^{\textrm{th}}$ & $\% 99^{\textrm{th}}$ & dev \\
         \hline\hline
         \multirow{4}{*}{2.4} & 0.6 & 0.9 & 1.08 & 1.39 & 1.44 & 7.7 \\ \cline{2-7}
         & \multirow{3}{*}{1.2} & 0.3 & 0.78 & 1.34 & 1.42 & 7.7 \\ \cline{3-7}
                              &                      & 0.6 & 1.08 & 1.63 & 1.71 & 7.8 \\ \cline{3-7}
                              &                      & 0.9 & 1.39 & 1.94 & 2.02 & 7.5 \\ \hline 
         3.6 & 1.8 & 0.3 & 1.11 & 1.90 & 2.01 & 7.5 \\ \hline
    \end{tabular}
\end{table}

We compute the difference between the word end time
from a GMM alignment and the chunk
end time in which the word is emitted by the chunked AED model.
Word emit latency measures can be found in \cref{tab:latency}.
Lookahead seems to add more latency compared to using larger chunk sizes (rows: 1 vs 2, 4 vs 5).

\subsection{Long-Form Recognition}

\begin{table}[t]
    \caption{
    \it WERs [\%] of \textbf{long-form speech recognition} on TED-LIUM-v2 test dataset with $\mathcal{C}$ concatenated sequences.
    }
    \label{tab:long-form-recog}
    \setlength{\tabcolsep}{2pt}
    \centering
    \begin{tabular}{|S[table-format=2]|c|c||S[table-format=2.1]|S[table-format=1.1]|S[table-format=1.1]|}
         \hline
         {\multirow{2}{*}{$\mathcal{C}$}} & \multicolumn{2}{c||}{Sequence lengths (sec)} & \multicolumn{2}{c|}{Global Enc.} & {Ch.~Enc.} \\ \cline{2-6}
         & \phantom{000.00}\llap{Mean} $\pm$ \rlap{Std}\phantom{00.00} & \phantom{00.00}\llap{Min} - \rlap{Max}\phantom{000.00} & {Gl.~Dec.} & \multicolumn{2}{c|}{Chunk Dec.} \\
         \hline\hline
         1 & \phantom{00}8.20 $\pm$ \phantom{0}4.30 & \phantom{0}0.35 - \phantom{0}32.55 & 6.9 & 6.9 & 7.3 \\ \hline
         2 & \phantom{0}23.10 $\pm$ \phantom{0}8.50 & \phantom{0}0.41 - \phantom{0}45.70 & 7.0 & 6.9 & 7.1\\ \hline
         4 & \phantom{0}33.70 $\pm$ 11.90 & \phantom{0}0.41 - \phantom{0}70.70 & 9.2 & 7.0 & 7.0 \\ \hline
         8 & \phantom{0}65.95 $\pm$ 22.19 & \phantom{0}7.19 - 116.99 & 23.4 & 7.1 & 7.1 \\ \hline
         10 & \phantom{0}82.51 $\pm$ 26.87 & 15.67 - 142.08 & 34.2 & 7.1 & 7.0 \\ \hline
         20 & 160.14 $\pm$ 53.98 & 17.83 - 237.27 & 62.4 & 7.1 & 7.0 \\ \hline
    \end{tabular}
\end{table}

We investigate the generalization on long-form speech recognition.
We conduct these experiments on TED-LIUM-v2 by
concatenating $\mathcal{C}$ consecutive sequences from the same recording to create much longer
sequences than what was seen in training.
We compare the global AED baseline to
a chunked AED model with left context carry-over $2.4$ sec, chunk size $1.2$ sec, lookahead $0.32$ sec
and
to a chunked-decoder with global encoder.
From the results in \cref{tab:long-form-recog},
we can observe that the global AED becomes much worse on longer sequences
whereas the chunked AED model generalizes very well and even improves, which is probably
because the decoder now has better LM context.
This is also the case when only the decoder is chunked.
The relative positional encoding in the encoder is probably helpful.
The generalization is much better than other variants such as segmental AED model \cite{zeyer2022:segmental-attention},
although that work uses an LSTM-based encoder.

\subsection{Beam Size and Length Normalization}

\begin{table}[t]
    \caption{
    \it Comparison of \textbf{effect of beam sizes and length normalization} between
    global AED and chunked AED models. WERs [\%] on TED-LIUM-v2 test dataset.
    }
    \label{tab:beam-size-len-norm}
    \centering
    \begin{tabular}{|c|S[table-format=2]||S[table-format=2.1]|S[table-format=1.1]|}
         \hline
         Length Norm. & {Beam} & {Global} & {Chunked} \\ \hline\hline
         (No influence) & 1    &  7.1 & 7.4 \\ \hline
         \multirow{3}{*}{Yes} & 12 & 6.9 & 7.3 \\ \cline{2-4}
                              & 32 & 6.9 & 7.3 \\ \cline{2-4}
                              & 64 & 6.9 & 7.3 \\ \hline
        \multirow{3}{*}{No}   & 12 & 7.0 &  7.3 \\ \cline{2-4}
                              & 32 & 8.5 & 7.3 \\ \cline{2-4}
                              & 64 & 10.9 & 7.3 \\ \hline
    \end{tabular}
\end{table}

Global AED model suffers from the length bias problem \cite{enc_dec_len_bias} 
because there is no explicit length modeling which pushes the model to prefer 
short hypothesis, especially when increasing the beam size.
However, the chunked AED model, like transducer, does not have the length
bias issue since this is modeled by the EOC symbol.
To verify this, we run experiments with different beam sizes and 
optional length normalization \cite{length_norm} for both global AED
and chunked AED model on TED-LIUM-v2 test dataset.
Results are shown in \cref{tab:beam-size-len-norm}.
The global AED model degrades a lot as we increase beam size and disable
length normalization whereas the chunked AED model does not need
such heuristic and performance remains consistent.
Additionally, both models perform marginally worse with greedy recognition.

\subsection{External Language Model}

\begin{table}[t]
    \caption{
    \it WERs [\%] with Transformer and LSTM \textbf{language model integration} on LibriSpeech dataset.
    }
    \label{tab:lm-fusion}
    \centering
    \begin{tabular}{|c|c|c||c|c|}
        \hline
        Model & LM & ILM & dev-other & test-other \\ \hline\hline
        \multirow{4}{*}{Global AED} & - & - & 5.6 & 5.7 \\ \cline{2-5}
                   & \multirow{2}{*}{LSTM} & No  & 4.6 & 5.0 \\ \cline{3-5}
                   &                       & \multirow{2}{*}{Yes} & 4.3 & 4.5 \\ \cline{2-2} \cline{4-5}
                   & Trafo                 &                      & 3.7 & 4.2 \\ \hline \hline
        \multirow{4}{*}{Chunked AED} & - & - & 6.2 & 6.2 \\ \cline{2-5}
                   & \multirow{2}{*}{LSTM} & No  & 5.2 & 5.3 \\ \cline{3-5}
                   &                       & \multirow{2}{*}{Yes} & 4.5 & 4.8 \\ \cline{2-2} \cline{4-5}
                   & Trafo                 &                      & 4.4 & 4.7 \\ \hline
    \end{tabular}
\end{table}

\Cref{tab:lm-fusion} shows results with LM integration on LibriSpeech dataset.
The chunked AED model used is the best model from \Cref{tab:chunked-enc-dec}.
Interestingly, the WER performance gap between global AED
and chunked AED is reduced when using LSTM LM and ILM subtraction.
Both models gain huge improvement from the LM integration.

\subsection{Comparison to Transducer}

\begin{table}[t!]
    \caption{
    \it WERs [\%] for \textbf{transition towards original transducer}, using global encoder, chunked decoder, chunk size 1.
    }
    \label{tab:transducer-variant}
    \setlength{\tabcolsep}{2pt}
    \centering
    \begin{tabular}{|l|c|c||c|c|}
        \hline
        \multirow{2}{*}{Model} & \multicolumn{2}{c||}{TED-v2} & \multicolumn{2}{c|}{LibriSpeech} \\ \cline{2-5}
                & dev & test & dev-other & test-other \\
        \hline\hline
        Baseline with $T'_w = 1$ & 7.5 & 7.3 & 5.8 & 6.0 \\ \hline
        \hspace{0.5mm} + EOC masking in $g$ & 7.6 & 7.2 & 5.8 & 6.1 \\ \hline
        \hspace{2.5mm} + Remove $c_s$ dep.~in $g$ & 7.7 & 7.4 & 6.0 & 6.1 \\ \hline
    \end{tabular}
\end{table}

We study the transition from a chunked AED model with chunk size 1 into a transducer model 
\cite{graves2012rnnt}.
The attention context vector $c_s$ in this variant is the encoder hidden representation $h_t$
at $t = k_s$ because the model attends to a single frame at a time.
We first mask out the $\eoc$ labels (blank in transducer) from the decoder LSTM $g$,
as the decoder LSTM in the original transducer only operates on non-blank labels.
Further, we completely remove the dependency to the encoder $h$ from the decoder LSTM,
just like in the original transducer.
Results are shown in \cref{tab:transducer-variant}.
We see that the additional dependencies seem to be helpful,
consistent with \cite{zeyer2020:transducer}.


\section{Conclusion}

In this work, we investigate a streamable chunked attention-based encoder-decoder (AED) model.
We show that this model is competitive compared to non-streamable global AED
model and generalizes very well on long-form speech recognition.
All degradations occur only in the chunked encoder
-- a chunked decoder with global encoder performs just as well as the global AED model.
We study the equivalence to the transducer model and find the extensions to be helpful.

\vspace{-1.5mm}
\begin{center}
\fontsize{6pt}{0pt}\selectfont
\bf ACKNOWLEDGEMENT
\end{center}
\vspace{-2mm}
{
\fontsize{5.5pt}{6pt}\selectfont
This work was partially supported by NeuroSys, which as part of the
initiative “Clusters4Future” is funded by the Federal Ministry of
Education and Research BMBF (03ZU1106DA), and by the project RESCALE
within the program \textit{AI Lighthouse Projects for the Environment,
Climate, Nature and Resources} funded by the Federal Ministry for the
Environment, Nature Conservation, Nuclear Safety and Consumer
Protection (BMUV), funding ID: 67KI32006A.
We thank Wei Zhou, Nick Rossenbach, Zoltán Tüske, Zijian Yang for useful discussions.

}


\bibliographystyle{IEEEbib-abbrev}

\let\OLDthebibliography\thebibliography
\renewcommand\thebibliography[1]{
  \OLDthebibliography{#1}
  \setlength{\parskip}{0pt}
  \setlength{\itemsep}{0.2pt}
  \setstretch{0.75}  
  \small
}

\bibliography{refs}

\end{document}